\documentclass[aam,referee]{svjour2}
\begin{document}

\title{Nonlocal description of a falling body through the air}
\author{Kwok Sau Fa }
\maketitle

\institute{Departamento de F\'{\i}sica, Universidade Estadual de Maring\'{a},\\
Av. Colombo 5790, 87020-900, Maring\'{a}-PR, Brazil, \ 
\email{kwok@dfi.uem.br}}

\abstract{
In this present work we consider a falling body through the air under the
influence of gravity. In particular, we consider the experimental data based
upon the free fall of \ six men in the atmosphere of the earth. In order to
describe this process we employ a nonlocal dissipative force. We show that
our description, by using an exponential memory kernel, can fit the
experimental data as well as that obtained by a local dissipative force.
}

\section{\protect\bigskip Introduction}

Recently, the idea of employing a nonlocal dissipative force has captured so
much interest in the scientific community. For instance, in diffusion
processes, generalized Langevin equations have been employed to describe
anomalous diffusion regimes, including both subdiffusion and superdiffusion.
According to Kubo \cite{kubo} the frictional memory kernel is related to the
form of the correlation function for random force when the system is in
equilibrium state. As can be noted, anomalous diffusion transport is present
in diverse physical systems, e.g., it can be found in porous media \cite%
{porous,porous1}, polymers \cite{polymers,polymers1}, amorphous
semiconductors \cite{semicond,semicond1} and composite heterogeneous films 
\cite{films}.

Approaches based on the fractional derivatives have also been employed to
describe, for instance, anomalous diffusion transport \cite{Metzler,Metzler1}
and viscoelastic systems \cite{bagley,makris,rossi}. The fractional
derivatives are also nonlocal and they can be related to memory effects.
Recent works on the theme, have revealed that a fractional derivative can
behave as a damping term \cite{Metzler,Metzler1}. In this respect, a simple
fractional oscillator has also been employed to investigate the physical
behavior of a fractional derivative, and it supports this point of view \cite%
{oscillator,oscillator1}.

In this work we consider a simple process involving a falling body through
the air. We also consider an observed process based upon the free fall of
six men in the atmosphere of the earth. As it is well-known the usual
description employs the Newton second law with a dissipative term in
function of the velocity of the falling body. For this observed process the
usual model, with the dissipative force proportional to $v^{2}$, \ is
employed. This phenomenological model is nonlinear. However, other
descriptions can also be used to describe the process and they can fit the
observed data equally well. In our description, we work on the assumption
that the dissipative force is nonlocal which is related to the memory
effect. We analyze our model by employing \ short and long memory kernels.
Our results are compared with the result of the usual approach and
experimental data.

\section{Local and nonlocal models}

When an object falls from rest through the air surrounding the earth, it
experiences a resisting force that opposes to the relative motion in which
the object moves relatively to the air. Experimentally, this resisting force
is related to the relative speed $v$. For slow speeds the resisting force is
in magnitude proportional to the speed. This last assumption is usually
employed to describe diffusion processes such as in the Langevin equation 
\cite{risken}. However, in other cases it may be proportional to the square
(or some other power) of the speed. Applying the Newton second law \cite%
{simon,spiegel} we have:we have: 

$$
m{{\hbox{d}v}\over {\hbox{d}t}}=mg-mF(v)\hbox{ ,}  \eqno (1)
$$
where $g$ is the free-fall acceleration and $F(v)$ is a function of the
velocity. We note that for $F(v)\sim v^{2}$, Eq. (1) is nonlinear
and it corresponds to the well-known \ Riccati equation \cite{Davis}. In
particular, for $F(v)=K_1 v/m$, we have the following solutions for $z$ and 
$v$:

$$
v(t)={mg \over K_1}-c_1 \exp \Bigl[ -{K_1 \over m}t\Bigr]  \eqno (2)
$$
and

$$
z(t)=z_0-{c_1 m \over K_1 }+{m \over K_1}\Bigl( gt+c_{1}\exp \Bigl[ -%
{K_1 \over m}t\Bigr] \Bigr) \hbox{ ,}  \eqno (3)
$$
where $c_1=mg/K_1-v_0$, $z_0$ is the initial position and $v_0$ is
the initial velocity. For $F(v)=K_2 v^{2}/m$, we obtain

$$
v(t)=b{1+c_{2}\exp \bigl[ -pt\bigr] \over 1-c_{2}\exp \bigl[ -pt\bigr] }
\eqno (4)
$$
and

$$
z(t)=z_{0}+{b \over p}\ln \biggl( {\bigl( 1-c_{2}\exp \bigl[ -pt\bigr]
\bigr) ^{2} \over (1-c_{2})^{2}\exp \bigl[ -pt\bigr] }\biggr) \hbox{ ,}
\eqno (5)
$$
where $b^{2}=mg/K_{2}$, $p=2bK_{2}/m$ and $c_{2}$=$\bigl( v_{0}-b\bigr)
/\bigl( v_{0}+b\bigr) $.

For these cases, Eq. (1) shows that the resisting force of an object
increases with the increase of the relative speed $v$. Consequently, when
the body falls enough, it reaches a constant terminal velocity. For
solutions (2) and (4) the terminal velocities are obtained
by taking $t\rightarrow \infty $, and the results are:

$$
V_{1}={mg \over K_1}  \eqno (6)
$$
and

$$
V_{2}=\sqrt{{mg \over K_2}}\hbox{ .}  \eqno (7)
$$

Now one considers the above system by assuming a nonlocal dissipative force
given by

$$
F(v)=\int_{0}^{t}\gamma \bigl( t-\tau \bigr) v\bigl( \tau \bigr) \hbox{d}%
\tau \hbox{ ,}  \eqno (8)
$$
where $\gamma (t)$ is the memory kernel. We note that Eq. (1) is
linear with $F(v)$ given by (8), and its solutions can be obtained
by using the Laplace transform. For $\gamma (t)\sim \delta (t)$ one has a
local dissipative force with $F(v)\sim v$. For a nonlocal memory kernel we
choose the exponential and power-law functions. We see that these functions
have short and long tails, respectively. Our idea, with these functions, is
to analyze the influence of memory function on the system described by Eq. (1) 
for short and long memory effects.

For exponential memory kernel $\gamma (t)=\gamma _{0}e^{-\lambda t}$ we
obtain the following solution:

$$
z=z_0+A+{g\lambda \over \gamma _{0}}t-\Bigl[ A-\Bigl( v_{0}-{%
g\lambda \over \gamma _{0}}-{\lambda \over 2}A\Bigr) t\Bigr] e^{-{%
\lambda \over 2}t}\hbox{, \ \ \ }\gamma _{0}={\lambda ^{2} \over 4}\hbox{ ,}
\eqno (9)
$$

$$
v(t)={g\lambda \over \gamma _{0}}+\Bigl[ v_{0}-{g\lambda \over \gamma _{0}}%
-{\lambda \over 2}\Bigl( v_{0}-{g\lambda \over \gamma _{0}}-{\lambda 
\over 2}A\Bigr) t\Bigr] e^{-{\lambda \over 2}t}\hbox{, \ }\gamma _{0}={%
\lambda ^{2} \over 4}\hbox{, }  \eqno (10)
$$
$$
z=z_{0}+A+{g\lambda  \over \gamma _{0}}t+ 
$$
$$
\Bigl[ \Bigl( v_{0}-{g\lambda \over \gamma _{0}}-{\lambda \over 2}A\Bigr) 
{\sin \bigl( B_{1}t\bigr) \over B_{1}}-A\cos \bigl( B_{1}t\bigr) \Bigr]
e^{-{\lambda \over 2}t}\hbox{, \ }\gamma _{0}>{\lambda ^{2}\over 4}\hbox{ ,%
}  \eqno (11)
$$

$$
v(t)={g\lambda \over \gamma _{0}}+ \Bigl( v_{0}-{g\lambda \over \gamma _{0}%
}\Bigr) \cos \bigl( B_{1}t\bigr) e^{-{\lambda \over 2}t}+ 
$$
$$
\Bigl[ AB_{1}-{\lambda \over 2B_{1}}\Bigl( v_{0}-{g\lambda \over \gamma
_{0}}-{\lambda \over 2}A\Bigr) \Bigr] \sin \bigl( B_{1}t\bigr) e^{-{%
\lambda \over 2}t}\hbox{, \ \ \ }\gamma _{0}> {\lambda ^{2}\over 4}\hbox{ , }
\eqno (12)
$$
$$
z=z_{0}+A+{g\lambda \over \gamma _{0}}t+ 
$$
$$
\Bigl[ \Bigl( v_{0}-{g\lambda \over \gamma _{0}}-{\lambda \over 2}A\Bigr) 
{\sinh \bigl( B_{2}t\bigr) \over B_{2}}-A\cosh \bigl( B_{2}t\bigr) \Bigr]
e^{-{\lambda \over 2}t}\hbox{, \ }\gamma _{0}< {\lambda ^{2} \over 4}\hbox{,}
\eqno (13)
$$
and

$$
v(t)={g\lambda \over \gamma _{0}}+ \Bigl( v_{0}-{g\lambda \over \gamma _{0}%
}\Bigr) \cosh \bigl( B_{2}t\bigr) e^{-{\lambda \over 2}t}- 
$$
$$
\Bigl[ AB_{2}+{\lambda \over 2B_{2}}\Bigl( v_{0}-{g\lambda \over \gamma
_{0}}-{\lambda \over 2}A\Bigr) \Bigr] \sinh \bigl( B_{2}t\bigl) e^{-%
{\lambda \over 2}t}\hbox{, \ \ \ }\gamma _{0}< {\lambda ^{2} \over 4}\hbox{
, }  \eqno (14)
$$
where $A=\bigl( g+\lambda v_{0}-g\lambda ^{2}/\gamma _{0}\bigr) /\gamma
_{0} $, $B_{1}=\sqrt{\gamma _{0}-\lambda ^{2}/4}$ and $B_{2}=\sqrt{\lambda
^{2}/4-\gamma _{0}}$.

For the three cases we can verify that $v(t=0)=v_{0}$ and they give the same
expression for the terminal velocity which is given by

$$
V={g\lambda \over \gamma _{0}}\hbox{.}  \eqno (15)
$$
Now, it is more convenient to write Eqs. (9), (11) and (13) in terms of the 
terminal velocity%
$$
z=z_{0}+{v_{0}V \over g}-{3 \over 4} {V^{2} \over g}+Vt+\Bigl[ {3 \over 4}%
{V^{2} \over g}+\Bigl( v_{0}+{V \over 2}\Bigr) t\Bigr] e^{-{2g \over V}t}%
\hbox{, \ \ \ }\gamma _{0}={\lambda ^{2} \over 4}\hbox{ ,}  \eqno (16)
$$
$$
z=z_{0}+A+Vt-\Bigl[ \Bigl( v_{0}-V-{\lambda \over 2}A\Bigr) {\sin
\bigl( B_{1}t\bigr) \over B_{1}}-A\cos \bigl( B_{1}t\bigr) \Bigr] e^{-{%
\lambda \over 2}t}\hbox{, \ }\gamma _{0}> {\lambda ^{2} \over 4}\hbox{,}
\eqno (17)
$$
and%
$$
z=z_{0}+A+Vt+ 
$$
$$
\Bigl[ \Bigl( v_{0}-V-{\lambda \over 2}A\Bigr) {\sinh \bigl(
B_{2}t\bigr) \over B_{2}}-A\cosh \bigl( B_{2}t\bigr) \Bigr] e^{-{\lambda 
\over 2}t}\hbox{, \ }\gamma _{0}< {\lambda ^{2}\over 4}\hbox{,}  \eqno (18)
$$
where $A=V/\lambda +v_{0}V/g-V^{2}/g$, $B_{1}=\sqrt{g\lambda /V-\lambda
^{2}/4}$ and $B_{2}=\sqrt{\lambda ^{2}/4-g\lambda /V}$. We note that $\gamma
_{0}$ does not appear explicitly in these last expressions. In particular,
the solution for $\gamma _{0}={\lambda ^{2} \over 4}$ only contains one
unknown parameter such as in the usual approach. For two other cases the
free fall can be described by one unknown parameter $\lambda $ which can be
determined by the experimental data.

In order to obtain some insights on the solutions of the nonlocal approach (9)-(14) 
and the solution of the usual approach given by
Eq. (5) let us fit the experimental data from the free fall of six
men in the atmosphere of the earth, from a maximum altitude of 31,400 to
2,100 feet which was accomplished in 116 seconds \cite{Davis}. The average
weight of the men and their equipment was 261.2 pounds and the average
velocity of the lowest point corresponding to a drop of 29,300 feet in 116
seconds is $V=251.4$ feet per second. The experimental data are shown in
Table 1 with the values of $H-z$, $H=31,400$.

To find the velocity of the lowest point $V$ we set $v_{0}=0$ and $z_{0}=0$.
We obtain from Eq. (5) the following approximate expression

$$
z=V_{2}t-\bigl( \ln 2\bigr) {V_{2}^{2} \over g}\hbox{ .}  \eqno (19)
$$
Substituting $z=29,300$, $t=116$ and $g=32.2$ we find $V_{2}=265.7$ ft/s.
This value is somewhat larger than the experimental result. In the case of $%
F(v)\sim v$, the terminal velocity is larger than $V_{2}$, thus this model
should be put aside. For the solution (16) we find the following
approximate expression%
$$
z=Vt-{3 \over 4} {V^{2} \over g}\hbox{ \ .}  \eqno (20)
$$
Note that the expressions (19) and (20) are similar and the
values $\ln 2$ and $3/4$ are close. From Eq. (20) we find \ $%
V=266.89$ ft/s. The parameter $\lambda $ can be estimated from (15)
and we find $\lambda =0.483$ s$^{-1}.$

In Fig. 1 we show the comparison of the observed values with those obtained
by the solutions (5) and (16). It is clear that the
solution (16) gives the adjustment as well as that of the usual
approach. In Fig. 2 we show the behavior of the solutions (5), (16), (17) and (18). 
For the solution (5) the terminal velocity is $V_{2}=265.7$ ft/s, while for other solutions the
terminal velocity is $V=266.89$ ft/s. The plots show that the curves are
close to each other. Thus, all of them can be used to describe the free fall
system equally well.

Next, we consider a long tail memory kernel given by $\gamma (t)=\gamma
_{1}t^{-\alpha }$ for $0<\alpha <1$. The solutions for $z(t)$ and $v(t)$ are
given by

$$
z(t)=z_{0}+gt^{2}E_{2-\alpha ,3}(-\gamma ^{\ast }t^{2-\alpha
})+v_{0}tE_{2-\alpha ,2}(-\gamma ^{\ast }t^{2-\alpha })  \eqno (21)
$$
and

$$
v(t)=gtE_{2-\alpha ,2}(-\gamma ^{\ast }t^{2-\alpha })+v_{0}E_{2-\alpha
,1}(-\gamma ^{\ast }t^{2-\alpha })\hbox{ ,}  \eqno (22)
$$
where $\gamma ^{\ast }=\gamma _{1}\Gamma \bigl( 1-\alpha \bigr) $, $\Gamma
(x)$ is the gamma function and $E_{\mu ,\nu }$ is the generalized
Mittag-Leffler function defined by $E_{\mu ,\nu }(z)=\sum_{n=0}^{\infty
}(z)^{n}/\Gamma (n\mu +\nu )$ \cite{Mainardi,erdel}. For $t=0$ we verify
that $v(t=0)=v_{0}$. In order to verify the terminal velocity we should
analyze the behavior of Eq.(22) for large time. To do so, we use the
following asymptotic approximation:%
$$
E_{\mu ,\nu }(z)\sim -{1 \over z\Gamma \bigl( \nu -\mu \bigr) }\hbox{ .}
\eqno (23)
$$
Then, we obtain from (22), for $v_0 =0$, the following
approximation:%
$$
v(t)\sim {gt^{\alpha -1} \over \gamma ^{\ast }\Gamma \bigl( \alpha \bigr) }%
\hbox{ .}  \eqno (24)
$$
This last result shows that Eq. (22) can not give a terminal
velocity. Then, a long time memory such as given by the power-law memory
kernel, based on Eqs. (1) and (8), is not appropriate to
describe the free fall system through the air.

\section{Conclusion}

In summary, we have investigated the falling body problem by using the local
and nonlocal dissipative forces. We note that nonlocal dissipative forces
are frequently used to describe diffusion processes by a generalized
Langevin equation. Then, this kind of approach represents a well-known
generalization of the usual approach. We have shown that the results of a
nonlocal dissipative force given by the exponential memory kernel can fit
the experimental data as well as the result obtained by the local
dissipative force $F(v)\sim v^{2}$. In contrast with the nonlinear force $%
F(v)\sim v^{2},$ the nonlocal dissipative force (8) gives a linear
differential equation. On the other hand, the power-law memory kernel does
not give a terminal velocity and consequently it alone can not be used to
describe the free fall system. These results suggest that a short memory
effect in the nonlocal approach may be appropriate to describe the free fall
system. The presence of memory effect in the system may be related to the
fact that when the body moves through the air, it displaces the particles in
its immediate vicinity. Then, the surrounding flow field is altered and acts
back on the body, resulting in an action which depends on the past motion of
the body. This feedback process which leads to a memory effect can be
reflected to the motion of a particle in a fluid \cite{lukic}. Further, we
have also verified Eq. (1) with an additional term for the
dissipative force given by $F(v)=K_{1}v/m+\gamma _{0}\int_{0}^{t}\exp \bigl(
-\lambda (t-\tau )\bigr) v\bigl( \tau \bigr) \hbox{d} \tau .$ For this case we
can improve the adjustment, i.e., we can fit the experimental data as well
as the result of the usual approach with a lower terminal velocity ($V=265$
ft/s). In Fig. 3 we show the comparison of the result with that described by
the usual approach. In the case of a fractional derivative, it can also be
employed to replace the classical Newtonian force, but it does not lead to
improve the result obtained by the usual approach significantly \cite{kwok}.

\bigskip

\begin{acknowledgements}
The author acknowledges partial financial support from the Conselho Nacional
de Desenvolvimento Cient\'{\i}fico e Tecnol\'{o}gico (CNPq), Brazilian
agency.
\end{acknowledgements}

\bigskip

\newpage

\bigskip

\begin{center}
{\Large Figure Captions}

\bigskip

\bigskip
\end{center}

\bigskip

Fig.1 - Adjustment of the experimental data with the expressions (5)
(dotted line) and (16) (solid line). The terminal velocity for the
expression (5) is given by $V_{2}=265.7$ ft/s, whereas the terminal
velocity for the expression (16) is given by $V=266.89$ ft/s.

\bigskip

Fig.2 - Plots of the expressions (5), (16), (17)
and (18). The terminal velocity for the expression (5) is
given by $V_{2}=265.7$ ft/s, whereas the terminal velocity for the three
other expressions is given by $V=266.89$ ft/s.

\bigskip

Fig. 3 - Adjustment of the experimental data with the expression (5)
(dotted line) and the result obtained by Eq. (1) with the
dissipative force given by $F(v)=K_{1}v/m+\gamma _{0}\int_{0}^{t}\exp \bigl(
-\lambda (t-\tau )\bigr) v\bigl( \tau \bigr) $d$\tau $ (solid line). The
terminal velocity for the expression (5) is given by $V_{2}=265.7$
ft/s, whereas the terminal velocity for the second case is given by $V=265$
ft/s, where $\lambda _{1}=\lambda +K_{1}/m$ and $\gamma _{0}+\lambda
K_{1}/m=\lambda _{1}^{2}/4$.

\bigskip

\newpage

\bigskip

\bigskip

\begin{tabular}{|ll||ll||ll|}
\hline
\ t & \ \ \ \ \ H-z & \ \ \ t & \ \ \ \ \ H-z & \ \ \ \ t & \ \ \ \ H-z \\ 
\hline
\ \ \ \ 0 & \ 31,400 & 46.7 & \ \ 20,550 & \ \ 88.1 & \ \ 9,400 \\ 
\ 9.9 & \ \ 30,780 & 51.3 & \ \ 19,400 & \ \ 92.7 & \ \ 8,140 \\ 
14.5 & \ \ 30,200 & 55.9 & \ \ 18,100 & \ \ 97.3 & \ \ 7,080 \\ 
19.1 & \ \ 28,850 & 60.5 & \ \ 16,600 & 101.9 & \ \ 5,700 \\ 
23.7 & \ \ 27,700 & 65.1 & \ \ 15,150 & 106.5 & \ \ 4,450 \\ 
28.3 & \ \ 26,150 & 69.7 & \ \ 14,070 & 111.1 & \ \ 3,170 \\ 
32.9 & \ \ 24,600 & 74.3 & \ \ 12,800 & 116.0 & \ \ 2,100 \\ 
37.5 & \ \ 23,200 & 78.9 & \ \ 11,650 &  &  \\ 
42.1 & \ \ 21,750 & 83.5 & \ \ 10,400 &  &  \\ \hline
\end{tabular}

\bigskip

\textbf{Table 1 }{\small Estimated values of }$\ H-z${\small \ for the fall
from an altitude of 31,400 to 2,100 feet, with }$z=0$ when $t=0${\small .}

\bigskip

\end{document}